\documentstyle[epsf]{elsart}

\begin{document}

\begin{frontmatter}

\title{Single massless Majorana fermion \\
  in the domain-wall formalism}

\author{Tomohiro Hotta}
\address{Institute of Physics, University of Tokyo, Komaba, Meguro-ku,
  Tokyo 153, Japan}
\author{Taku Izubuchi}
\address{Institute of Physics, University of Tsukuba, Tsukuba, Ibaraki
  305, Japan}
\author{Jun Nishimura}
\address{Department of Physics, Nagoya University, Chikusa-ku, Nagoya
  464-01, Japan}

\begin{abstract}
  We study the domain-wall formalism with additional Majorana mass
  term for the unwanted zero mode, which has recently been proposed
  for lattice construction of 4D ${\cal N}=1$ super Yang-Mills theory
  without fine-tuning.
  Switching off the gauge field, we study the dispersion relation of
  the energy eigenstates numerically, and find that the method works
  for reasonable values of Majorana mass.
  We point out, however, that a problem arises for too large Majorana
  mass, which can be understood in terms of the seesaw mechanism.
\end{abstract}


\end{frontmatter}

\section{Introduction}

There has been a remarkable progress in understanding non-perturbative
aspects of supersymmetric gauge theories recently.
The exact results for 4D ${\cal N} = 1$ and ${\cal N}=2$
supersymmetric gauge theories have been obtained using peculiar
properties of those theories such as the non-renormalization theorem,
the exact $\beta$ functions and so on.
The analytic progress is, however, naturally restricted to the aspects
of the theories which are strongly constrained by the supersymmetry,
and the properties of the non-BPS states, for example, are not
understood at all.
From this point of view, lattice study of these theories would
complement the recent progress and would provide us with a new insight 
into their nonperturbative dynamics.

Unfortunately supersymmetry is difficult to realize on the lattice.
This is not so surprising since the lattice regularization breaks the
translational invariance, which forms a subgroup of the
supersymmetry.
As the translational invariance is restored in the continuum limit, we 
can restore supersymmetry in the continuum limit.
But the price we have to pay for the latter is that we need
fine-tuning in general.
For 4D ${\cal N}=1$ supersymmetric Yang-Mills theory, one can use the
Wilson-Majorana fermion for the gaugino and recover supersymmetry in
the continuum limit by fine-tuning the hopping parameter to the chiral
limit \cite{CV}.
Some numerical works have been started along this line \cite{num}.
Fine-tuning is a hard task, however, as is known in the numerical
studies of the chiral limit in QCD, and a method without fine-tuning
is highly desired.

The overlap formalism \cite{NN} can be used for this purpose, since it 
preserves exact chiral symmetry on the lattice.
The problem here is that the formalism is not suitable for numerical
simulation as it stands.
A practical proposal%
\footnote{While this work was being completed, a preprint \cite{Neu}
  appeared which includes an alternative proposal in this direction.}
made by Ref. \cite{nishimura} is to use the domain-wall formalism
\cite{kaplan,FS}, and to decouple the unwanted zero mode by adding
Majorana mass term for it.
We examine whether this approach really works when the gauge field is
switched off as a first step.
We study the dispersion relation for various values of the Majorana
mass, and examine whether the model gives the desired spectrum.
We confirm that the model is fine for moderate values of the Majorana
mass, while for too large Majorana mass, an extra almost massless mode 
appears, which we must be careful of in future study of this model in
a more realistic situation with dynamical gauge field.

The paper is organized as follows.
In Sec.~\ref{model} we define our model and review the idea to obtain
single massless Majorana fermion without fine-tuning.
In Sec.~\ref{hamiltonian} we give the explicit form of the Hamiltonian
of the system.
We diagonalize it numerically to study the dispersion relation of the
energy eigenstates for various values of the additional Majorana mass.
In Sec.~\ref{interpretation} we interpret the appearance of the extra
massless mode for too large Majorana mass in terms of the seesaw
mechanism.
Section~\ref{summary} is devoted to summary and discussions.

\section{The model}
\label{model}

Four-dimensional ${\cal N} = 1$ super Yang-Mills theory contains the
gauge boson and the gaugino.
The gaugino is Majorana fermion, which is equivalent to Weyl fermion
in four dimensions.

In order to avoid fine-tuning, we need to impose the chiral symmetry
on the lattice.
This can be done by the domain-wall formalism \cite{FS}.
Here we have two Weyl fermions with opposite handedness, which couple
to the gauge field in the vector-like way.

The idea of Ref. \cite{nishimura} is to apply this formalism to 4D
super Yang-Mills theory, by decoupling one of the Weyl fermions by
giving it mass of the order of the cutoff through the additional
Majorana mass term for it.
It should be noted that this can be done without violating the gauge
invariance, since the fermion is in the adjoint representation, which
is a real representation, for which the Majorana mass term in 4D is
gauge invariant.
As a first step, we examine this model by switching off the gauge
field.

The action of the model consists of two parts:
\begin{equation}
  \label{action}
  S = S_0 + S_{\rm mass}.
\end{equation}
$S_0$ is given by
\begin{equation}
  \label{action0}
  S_0 = \bar{\xi} \sigma_\mu \partial_\mu \xi + \bar{\eta}
  \bar{\sigma}_\mu \partial_\mu \eta 
  + \bar{\xi} {\cal M} \eta +\bar{\eta} {\cal M}^\dagger \xi, 
\end{equation}
where
\begin{equation}
  {\cal M} = \partial_s + M + \frac{1}{2} \triangle. 
\end{equation}

$\xi(x_\mu, s)$ and $\eta(x_\mu, s)$ are right-handed and left-handed
Weyl fermions respectively in the four-dimensional space-time, which
is latticized as $\{ x_\mu \in {\cal Z};\mu=1,2,3,4 \}$.
The $s$ denotes the coordinate in the fifth direction, which runs over
$1, \cdots, N_s$.
The boundary condition is taken to be free in the fifth direction, and
to be periodic in the four space-time directions.
Summation over the five-dimensional coordinates ($x_\mu,s$) is
suppressed in Eq. (\ref{action0}) and similar abbreviations are used
in the rest of this paper.
$\triangle$ is the five-dimensional lattice Laplacian.
$\partial$ should be understood as the lattice derivative.
$\sigma_\mu$ and $\bar{\sigma}_\mu$ are defined by $\sigma_\mu = (1, i
\sigma_i)$ and $\bar{\sigma}_\mu = (1, -i \sigma_i)$, where $\sigma_i$
($i=1,2,3$) are the Pauli matrices.
$M$ is a mass parameter, which is fixed at some value within $0 < M <
1$, when one takes the continuum limit.

With the action $S_0$, one obtains a right-handed Weyl fermion and a
left-handed one localized at the boundaries of the fifth direction $s
= 1, N_s$, respectively.
In the $N_s \rightarrow \infty$ limit, the chiral symmetry is exact
and we end up with one massless Dirac fermion \cite{FS}.
For finite $N_s$, the chiral symmetry is violated, but the violation
vanishes exponentially with increasing $N_s$ \cite{Neu,Vranas}.

Let us identify the zero mode in $\xi$ as the massless Majorana
fermion we want, namely the gaugino.
In order to decouple the unwanted zero mode in $\eta$, we introduce
the additional term $S_{\rm mass}$ in the action.
There is a variety of choice for the $S_{\rm mass}$.
We can, for example, take the Majorana mass term given by 
\cite{nishimura}.
\begin{equation}
  \label{majorana}
  S_{\rm mass} = \displaystyle m \left. \left ( \eta^T_s \sigma_2
  \eta_s + \bar{\eta}_s \sigma_2 \bar{\eta}^T_s \right ) \right|_{s =
    N_s}.
\end{equation}
$m$ is the parameter which we refer to as the Majorana mass.
It should be kept fixed, when one takes the continuum limit, in order
to give mass of the order of the cutoff to the unwanted zero mode.

\section{The dispersion relation}
\label{hamiltonian}

We examine the dispersion relation to see if we get the desired
spectrum.
This analysis has been done for the domain-wall formalism without the
extra Majorana mass term in Ref. \cite{jansen}.

The Hamiltonian of the system can be obtained from the action
(\ref{action}) as
\begin{eqnarray}
  \label{eqn:hamiltonian}
  H & = & -i \xi^\dagger \sigma_i \partial_i \xi
  +i \eta^\dagger \sigma_i \partial_i \eta
  + \xi^\dagger \partial_s \eta - \eta^\dagger \partial_s \xi
  -M (\xi^\dagger \eta + \eta^\dagger \xi)  \nonumber \\
  & & - \frac{1}{2} ( \xi^\dagger \triangle_4 \eta
  + \eta^\dagger \triangle_4 \xi )
  - m \left. \left ( \eta^T \sigma_2 \eta + \bar{\eta} \sigma_2
      \bar{\eta}^T \right ) \right|_{s = N_s} ,
\end{eqnarray}
where $\triangle_4$ represents the lattice Laplacian in the $(x,y,z,s)$
directions.
Since we have switched off the gauge field, the system is
translationally invariant in the $(x,y,z)$ directions, and therefore
we can partially diagonalize the Hamiltonian by working in the
momentum basis.

The Hamiltonian for each three-dimensional momentum $\bf p$ can be
given as
\begin{eqnarray}
  H({\bf p}) & = & \xi^\dagger({\bf p}) \sigma_i \sin p_i \, \xi({\bf
    p}) 
  - \eta^\dagger({\bf p}) \sigma_i \sin p_i  \, \eta({\bf p})
  \nonumber \\
  & & + \xi^\dagger({\bf p}) \left\{ \partial_s - \frac{1}{2}
    \triangle_s - M - (\cos p_i - 1) \right\} \eta({\bf p}) \nonumber
    \\
  & & + \eta^\dagger({\bf p}) \left\{ - \partial_s - \frac{1}{2}
    \triangle_s - M - (\cos p_i - 1) \right\} \xi({\bf p}) \nonumber
  \\
  & & + [{\bf p} \leftrightarrow - {\bf p} ] \nonumber \\
  & & - 2m \left. \left\{ \eta^T(-{\bf p}) \sigma_2 \eta({\bf p})
      + \eta^\dagger({\bf p}) \sigma_2 \eta^\ast(-{\bf p})
    \right\} \right|_{s = N_s} , 
\end{eqnarray}
where $\triangle_s$ is the lattice Laplacian in the $s$ direction.
We note that the Majorana mass term mixes the fields with the momentum
$\bf p$ and those with $\bf -p$.

We calculate numerically the eigenvalues of the above Hamiltonian for
each three-dimensional momentum $\bf p$ for various values of the
Majorana mass term.
The only difference from the analysis in Ref. \cite{jansen} is that
since the fermion number is not conserved due to the Majorana mass
term, we have to make a Bogoliubov transformation to diagonalize the
Hamiltonian.

\begin{figure}[htbp]
  \begin{center}
    \leavevmode
    \epsfysize=7cm
    \epsffile{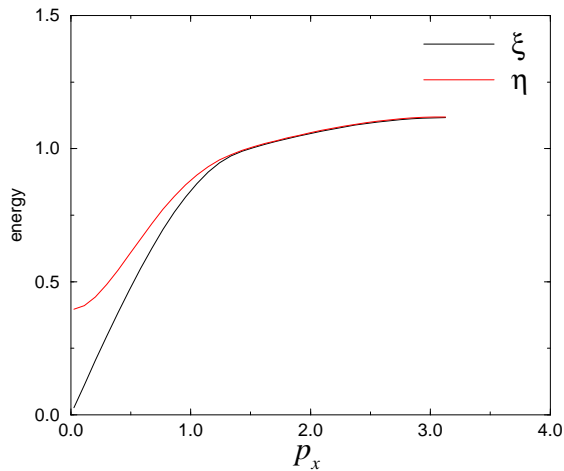}
    \caption{The dispersion relation of $\xi$ and $\eta$ for $m =
      0.2$.}
    \label{fig:dispersion1}
  \end{center}
\end{figure}

For moderate Majorana mass we have one massless Weyl fermion localized
at $s = 1$ as expected.
Figure~\ref{fig:dispersion1} shows the energy of $\xi$ and $\eta$ as a 
function of $p_x$ with $p_y = p_z = 0$ when the Majorana mass is
$0.2$. Here and henceforth, we take $N_s = 20$ and $M = 0.9$.
One can see that the $\xi$ has a linear dispersion relation near the
origin $p_x = 0$, while the $\eta$ has a mass gap of order one.
The doublers of $\xi$ and $\eta$ are removed as in the case without
the Majorana mass term.

\begin{figure}[htbp]
  \begin{center}
    \leavevmode
    \epsfysize=7cm
    \epsffile{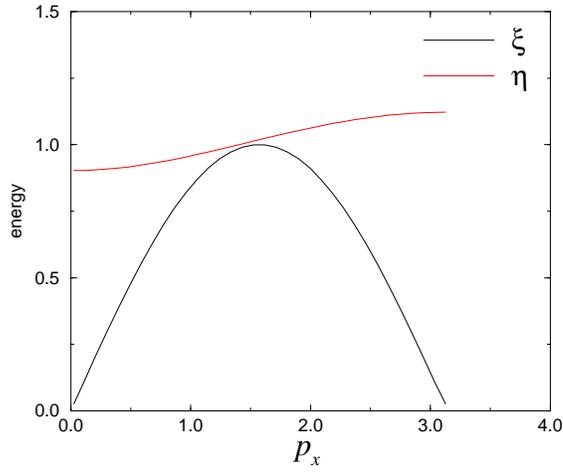}
    \caption{The dispersion relation of $\xi$ and $\eta$ for $m =
      1000$.}
    \label{fig:dispersion2}
  \end{center}
\end{figure}

One might think that larger Majorana mass only results in larger mass
for $\eta$ without any problem, but this is not the case.
Figure~\ref{fig:dispersion2} shows the dispersion relation for large
Majorana mass $m = 1000$. 
One can see that although the $\xi$ remains massless and the $\eta$
massive, the doublers of $\xi$ become very light.

\begin{figure}[htbp]
  \begin{center}
    \leavevmode
    \epsfysize=7.5cm
    \epsffile{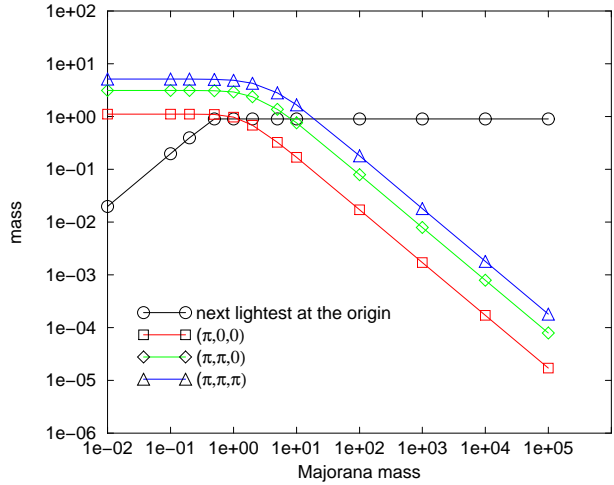}
    \caption{The mass of the next lightest mode at $\bf p = \bf 0$
      as well as that of the doublers of $\xi$ as a function of the
      Majorana mass.} 
    \label{fig:doubler}
  \end{center}
\end{figure}
In Fig.~\ref{fig:doubler} we plot the mass of the next lightest mode
at $\bf p = \bf 0$ as well as that of the doublers of $\xi$ as a
function of the Majorana mass.
The doublers have mass of the order of the cutoff for $m <1.0$, while
for $m > 1.0$, the mass decreases with increasing Majorana mass as
$\sim 1/m$.
The doublers with many $\pi$'s in the momentum components are heavier
than those with less $\pi$'s.
The mass of the next lightest mode at $\bf p = \bf 0$ grows linearly
as the Majorana mass increases, but saturates for $m>0.5$.

The results in this section lead us to the conclusion that there is an 
appropriate range of the Majorana mass to obtain single Majorana
fermion.
Note, however, that this does not mean the need for fine-tuning of the
parameter since we have quite a large allowed range for the Majorana
mass.

\section{Interpretation of the result for the large Majorana mass}
\label{interpretation}

We first confirm the behavior of the doubler mass in the large
Majorana mass case by looking at the poles of the propagator, which
give the masses of the intermediate states.
The propagator of the fermions in the domain-wall formalism with the
Majorana-type coupling has been calculated in Ref. \cite{aoki}.
The one for $\eta$ can be written as
\begin{equation}
  < \eta_s \bar{\eta}_t > = - \sigma_\mu \partial_\mu
  \left \{ A_R e^{-\alpha (s + t)} + A^m_L e^{\alpha (s + t - 2N_s)}
    + B e^{-\alpha |s - t|} \right \} ,
\end{equation}
where $A_R$, $A^m_L$ and $B$ are functions of the external momentum
$p$.
$\alpha$ is a positive constant determined by the parameters in the
action.
In Ref. \cite{aoki}, they examined $A_R$, $A^m_L$ and $B$ in the 
$p \rightarrow 0$ limit 
and showed that there exists no pole at $p^2 = 0$ when the Majorana
mass is non-zero, which means that the $\eta$ has been made heavy
successfully.

Similarly we can see the existence of very light doublers for the
large $m$, by looking at the behavior of the propagator of $\xi$ when
$p$ is near one of the corners of the Brillouin zone.
We extract the masses of the almost massless doublers from the singular
part of the propagator of $\xi$ as
\begin{equation}
  \label{mdoubler}
  m_{\rm doubler} \sim \frac{(2n - M + 2)(2n - M)}{m},
\end{equation}
for $m \gg 1$, where $n$ is the number of $\pi$'s in the momentum
components of the doubler.
We have checked that the masses of the doublers extracted from the
Hamiltonian diagonalization as in the previous section fit exactly to
this formula.

We note that the behavior of the doublers for large Majorana mass
seen above can be understood intuitively in terms of the seesaw
mechanism, which was originally proposed to explain the lightness of
neutrino.
A typical example of the mechanism is given by the case in which Dirac
and Majorana mass terms coexist.
When we diagonalize the mass matrix of the fermion, a very small
eigenvalue appears when the Majorana mass is much larger than the
Dirac mass.

In fact, the doublers have the two types of mass term in our model.
The Dirac mass term comes from the Wilson term in (\ref{action0}) and
can be written as
\begin{equation}
  S_{\rm Dirac} = 2 n  ( \bar{\xi} \eta + \bar{\eta} \xi ),
\end{equation}
where $n$ is the number of $\pi$'s in the momentum components of the
doubler as before.
Together with the Majorana mass term which comes from
(\ref{majorana}), we have the following mass matrix for the doublers
in the basis of the two-component Weyl fermion.
\begin{equation}
  \left(
    \begin{array}{cc}
      0 & 2n \\
      2n & m
    \end{array}
  \right).
\end{equation}
The eigenvalues $\lambda$ of this matrix for $m \gg n$ are given by
\begin{equation}
  \label{eigenvalue}
  \lambda \simeq m , \ \frac{4n^2}{m}.
\end{equation}
The second one reproduces the large $m$ behavior of the masses of
the doublers.

\section{Summary and Discussion}
\label{summary}

We examined whether the proposal for decoupling the unwanted zero-mode 
in the domain-wall approach by adding the Majorana mass term for it
works when the gauge field is switched off.
Above all, we clarified what values we should take for the Majorana
mass to be added.
We observed the desired dispersion relation for moderate values of the
Majorana mass, which means that the approach is promising.
We pointed out, however, that for too large Majorana mass, the
doublers of the desired Majorana fermion become very light.
We gave a natural explanation of this phenomenon in terms of the
seesaw mechanism.
We also confirmed our conclusion by the analysis of the fermion
propagator.

There are various types of additional Majorana mass term $S_{\rm
  mass}$ that can be used instead of the particular one we used above.
We checked that the following alternatives can be used successfully to
give $\eta$ mass of the order of the cutoff, while keeping $\xi$
massless.

\begin{enumerate}
\item Majorana mass term for both $\xi$ and $\eta$ localized at $s =
  N_s$.
  
\item Majorana mass term for $\eta$ in some finite region near
  $s=N_s$.
  One can even extend the region to cover the whole extent of the
  fifth direction.

\item Majorana mass term for both $\xi$ and $\eta$ in some finite
  region near $s=N_s$.
  Unlike the case (ii), one {\it cannot} extend the region to cover
  the whole extent of the fifth direction in this case.
\end{enumerate}

In either case, unwanted light modes appear when we take the Majorana
mass too large.

Our next task is of course to switch on the gauge field.
We also have to introduce additional boson fields to subtract the
heavy modes in the bulk as in Ref. \cite{Neu,Vranas}.
We hope this approach will finally enable us to understand general
nonperturbative phenomena in the super Yang-Mills theory, including
the ones related to the vacuum structure such as gaugino condensation.

\section*{Acknowledgements}

The authors would like to thank S.~Aoki, K.~Nagai and S.V.~Zenkin for 
discussions.
T.H. and T.I. are JSPS Research Fellow.

\end{document}